\documentclass[sigconf,screen,authorversion,nonacm]{acmart}

\usepackage{amsmath}
\usepackage{algorithm2e}
\usepackage{algorithmic}
\usepackage{alias}
\usepackage{newfloat}
\usepackage{listings}
\usepackage{enumitem}
\usepackage{bm}

\AtBeginDocument{%
  \providecommand\BibTeX{{%
    \normalfont B\kern-0.5em{\scshape i\kern-0.25em b}\kern-0.8em\TeX}}}

\acmConference[]{}{,
  }{}
\acmPrice{15.00}
\acmISBN{978-1-4503-XXXX-X/18/06}

\begin{document}

\title{Probabilistic Rank and Reward: A Scalable Model for Slate Recommendation}

\author{Imad Aouali}
\authornote{Both authors contributed equally to this research.}
\affiliation{%
  \institution{Criteo AI Lab and CREST ENSAE}
  \country{Paris, France}
}

\author{Achraf Ait Sidi Hammou}
\authornotemark[1]
\affiliation{%
  \institution{Criteo AI Lab}
    \country{Paris, France}
}

\author{Otmane Sakhi}
\affiliation{%
  \institution{Criteo AI Lab}
    \country{Paris, France}
}

\author{David Rohde}
\affiliation{%
  \institution{Criteo AI Lab}
    \country{Paris, France}
}

\author{Flavian Vasile}
\affiliation{%
  \institution{Criteo AI Lab}
    \country{Paris, France}
}

\renewcommand{\shortauthors}{Aouali and Ait Sidi Hammou, et al.}

\begin{abstract}
We introduce \textbf{P}robabilistic \textbf{R}ank and \textbf{R}eward (\textbf{PRR}), a scalable probabilistic model for personalized slate recommendation. Our approach allows off-policy estimation of the reward in the scenario where the user interacts with at most one item from a slate of $K$ items. We show that the probability of a slate being successful can be learned efficiently by combining \emph{the reward}, whether the user successfully interacted with the slate, and \emph{the rank}, the item that was selected within the slate. \textbf{PRR} outperforms existing off-policy reward optimizing methods and is far more scalable to large action spaces. Moreover, \textbf{PRR} allows fast delivery of recommendations powered by maximum inner product search (MIPS), making it suitable in low latency domains such as computational advertising.
\end{abstract}





\maketitle

\section{Introduction} \label{sec:introduction}

Recommender systems (advertising, search, music streaming, etc.) are becoming prevalent in society helping users navigate enormous catalogs of items to identify those relevant to their interests. In practice, recommender systems must optimize the content of an entire section of the web page that the user is navigating. This section is viewed as an ordered collection (or slate) of $K$ items \citep{swaminathan2017offpolicy,chen2019top,aouali2021combining}. It often takes the form of a menu and the user can choose to interact with one of its items. Both in academia and industry, A/B tests are seen as the golden standard to measure the performance of recommender systems. A/B tests enable us to directly measure utility metrics that rely on interventions, being the slates shown to the user. However, they are costly. Thus a clear need remains for reliable offline procedures to propose candidate recommender systems and consequently reduce the cost of A/B tests. 

In this work, we propose a probabilistic model called \textbf{P}robabilistic \textbf{R}ank and \textbf{R}eward (\textbf{PRR}) for large-scale slate recommendation. 
\textbf{PRR} addresses the following practical limitations of existing methods. \begin{enumerate}[label=\textbf{\arabic*)}]
    \item Collaborative filtering and content-based recommender systems \citep{su2009survey,lops2011content} optimize \emph{proxy} metrics of the reward. This may lead to a striking gap between their offline evaluation and the A/B test result \citep{Garcin2014}.
    \item Counterfactual estimators, which are often based on inverse propensity scoring (IPS) \citep{horvitz1952generalization}, suffer high bias and variance \citep{swaminathan2017offpolicy} in large-scale scenarios. Moreover, policy learning objectives for these estimators are mostly not suitable for slates and large action spaces.
    \item The decision rules produced by most existing methods do not fit the engineering constraints for deployment in large-scale, low-latency systems such as computational advertising; they are either expensive or intractable. To address these challenges, our paper makes the following contributions.
    \end{enumerate}

\textbf{1. Problem formulation:} We formalize the ubiquitous slate recommendation setting where the user is shown a slate of $K$ items and they can choose to interact with \emph{at most} one of its items. After that, the feedback consists of two signals: did the user successfully interact with one of the items? Then if an item was interacted with, which one was it? These are referred to by \emph{reward} and \emph{rank}, respectively. Note that it is very common in practice that the user interacts with \emph{at most} one item in the slate. For example, in ad placement, a click on an item causes the whole slate to disappear. As a result, the user cannot click on the other items in that slate. Similarly, in video recommendation, a click on a video to watch it reloads the homepage and changes it. Most offline reward optimizing methods do not consider this and assume that the user can interact with multiple items in the slate.

\textbf{2. Modeling the reward and rank:} We propose a probabilistic model (\textbf{PRR}) that combines both signals, the reward and rank. This is important as both contain useful information about the user interests and discarding one of them may lead to inferior performance. Existing methods either use one signal or partially combine the two by assuming that the reward is a function of the rank and only use the latter.

\textbf{3. Incorporating extra features:} \textbf{PRR} distinguishes between slate and item level features that contribute to an interaction with the slate and one of its items, respectively. \textbf{PRR} also incorporates that interactions can be predicted by \emph{engagement features} that neither represent the user interests nor the recommended items. This includes the slate size and the overall level of user activity and engagement. While these features are not used in decision making, incorporating them helps learn the user interests more accurately. Precisely, it allows differentiating between interactions that are caused by the quality of recommendations and those that happen due to overall user engagement. While the use of nuisance parameters to enhance estimation quality is not novel, the distinctiveness of our approach lies in the separation between user interest and engagement features, coupled with their application to our specific setting described above.

\textbf{4. Fast decision making:} \textbf{PRR}'s decision rule is reduced to solving a maximum inner product search (MIPS). This allows fast delivery of recommendations in large-scale tasks using approximate MIPS algorithms \citep{shrivastava2014asymmetric}. Most existing reward optimizing methods do not take into account this practical consideration and as such propose expensive decision rules. 

\textbf{5. Experiments:} We show empirically that \textbf{PRR} outperforms commonly used baselines in terms of both empirical performance and computational efficiency.

\section{Related Work}
\label{sec:sota}
A significant part of the classic recommender systems literature is inspired by the Netflix prize \citep{bennett2007netflix} which formulated recommendation as predicting item ratings in a matrix. A practitioner will often work with rating datasets that include neither recommendations nor rewards, but are suitable for collaborative filtering \citep{su2009survey} or content-based recommendation \citep{lops2011content}. While interesting, these datasets and problem formulations do not reflect the actual interactions between the users and the recommender systems; they are only imperfect proxies. In particular, the performance of an algorithm on these problems and datasets may be very different from its actual A/B test performance \citep[Section~5.1]{Garcin2014}. Instead, \emph{off-policy, or offline, reward optimizing recommendation} approaches aim at directly optimizing the reward using \emph{logged data} summarizing the previous interactions of users with the existing recommender system. These are also different from on-policy, or online, approaches that we do not cover in this work. Here the reward is learned offline using logged data.

\paragraph{Inverse Propensity Scoring (IPS)} Here we assume that the recommender system is represented by a stochastic policy $\pi$. That is, given a user $u$, $\pi(\cdot \mid u)$ is a probability distribution over the set of items. Under this assumption, \citet{dudik2014doubly} used inverse propensity scoring (IPS) \citep{horvitz1952generalization} to estimate the reward for recommendation tasks with small action spaces. Unfortunately, IPS can suffer high bias and variance in realistic settings such as slate recommendation. The high variance of IPS is acknowledged and several fixes have been proposed such as clipping, and self-normalization \citep{gilotte2018offline}. Another solution is doubly robust (DR) \citep{dudik2014doubly} which combines a reward model with IPS to reduce the variance. DR relies on a reward model and \textbf{PRR} can be integrated into it. 

In slate recommendation, recent works made simplifying structural assumptions to reduce the variance. For instance, \citet{li2018offline} restricted the search space by assuming that items contribute to the reward individually. Similarly, \citet{swaminathan2017offpolicy} assumed that reward is additive w.r.t. unobserved and independent ranks. The independence assumption is restrictive and can be violated in many production settings. A relaxed assumption was proposed in \citet{mcinerney2020counterfactual} where the interaction with the $\ell$-th item in the slate, $s_\ell$, depends only on $s_\ell$, $s_{\ell-1}$ and its rank $r_{\ell-1}$. This sequential dependence scheme is not sufficient to encode our setting where the user views the whole slate at once and interacts with \emph{at most} one of its items. \textbf{PRR} is model-based as it does not use inverse propensity scoring. 

\paragraph{Direct Methods (DM)} Here a reward model is learned and then used to estimate the performance of the recommender system. Existing methods \citep{sakhi2020blob,jeunen2021pessimistic} focus on single-item recommendation (slates of size 1) and do not incorporate engagement features. Another popular family related to direct methods is called \textit{click (or ranking) models} \citep{chuklin2015click}. Click models are often represented as graphical models and as such define dependencies manually and are not always scalable to large action spaces. Moreover, they do not incorporate extra features that are available in recommendation since they were primarily designed for search engine retrieval. Recently, \citet{cief2022pessimistic} used direct methods with pessimism in learning to rank. However, the models used in their work are not suitable for our scenario where the user is presented with the entire slate simultaneously and can interact with, at most, one item. To be precise, the dependent-click and position-based models described in \citet{cief2022pessimistic} allow multiple clicks on the slate, making them unsuitable for our setting. Moreover, the cascading models in \citet{cief2022pessimistic,kiyohara2022doubly} assume that the user interacts with items sequentially. That is, item $s_\ell$ at position $\ell$ is examined only if item $s_{\ell-1}$ at the previous position is examined but not clicked. This differs from our setting, where all items are examined simultaneously (rather than sequentially) and the user can click on, at most, one of them.  

\paragraph{Policy Learning} Finding an optimal policy to implement in a recommender system requires both estimating the expected reward of policies (using either direct or IPS methods above) and then searching the space of policies to find the one with the highest expected reward (policy learning). The literature mainly focuses on combining IPS with a softmax policy for single item recommendation.  Extending this to slates is challenging. In fact, it is tempting to use factored softmax policies but this may cause the learned policy to recommend slates with repeated items. This was acknowledged in \citet{chen2019top} that introduced a top-K heuristic to prevent the collapse of the policy on a single item. The other alternative is to model the policies using a Plackett-Luce distribution. Then, the straightforward approach to solving the resulting optimization problem involves using REINFORCE-style algorithms, which estimate gradients by sampling a slate from the policy. However, naive REINFORCE can be computationally expensive due to the $\mathcal{O}(P)$ cost of sampling from the policy and the high level of noise in the gradient estimates. Therefore, recent papers \citep{oosterhuis2021computationally,oosterhuis2022learning} proposed methods to accelerate the optimization process. Precisely, \citet{oosterhuis2021computationally,oosterhuis2022learning} introduced lower variance gradient estimators by assuming certain restrictive linear assumptions about the reward function. Although this estimator is less noisy, the iteration cost of the algorithm remains $\mathcal{O}(P)$. However, convergence with fewer iterations can be achieved in stochastic gradient descent (SGD) due to the lower variance. In our case, policy learning is not needed as the optimal policy of \textbf{PRR} falls out neatly to a MIPS task due to its parametrization. Also, the training time scales with the slate size $K$ rather than the action space $P$ as we show later.  

\paragraph{Decision Making} An important practical challenge is the design of tractable decision rules that satisfy engineering constraints. In practice, we must quickly recover a slate of items $\bm{s}$ given a context vector $\bm{x}$. The optimal decision rule under either the IPS or DM 
formulation amounts to solving ${\rm argmax}_{\bm{s} \in \mathcal{S}} L(\bm{x}, \bm{s})$. Here $L(\bm{x}, \bm{s})$ might either be the reward estimate in DM or the policy in IPS. However, the space of eligible slates $\mathcal{S}$ is combinatorially large. Thus exhaustively searching the space of slates is untenable and we must resort to finding good but implementable decision rules rather than optimal ones. There are three main strategies for doing this.  \begin{enumerate}[label=\textbf{(\Alph*)}]
    \item Reducing the combinatorial search over $\mathcal{S}$ to a sorting operation over a set of $P$ items.
    \item Two-stage systems that massively reduce candidate sets with a pre-filtering operation.
    \item Fast approximate maximum inner product search (MIPS) on which we focus in this work since it allows end-to-end optimization as opposed to the two-stage scheme. Next, we provide more detail for each solution.
    \end{enumerate}
    
\textbf{(A) Reducing the combinatorial search to sorting:} To accelerate decision making, existing methods moved the search space from the combinatorially large set of slates $\mathcal{S}$ to the catalog of items $\{1, \ldots, P\}$. This is achieved by first associating a score for items instead of slates and then recommending the slate composed of the top-K items with the highest scores. This leads to a $\mathcal{O}(P)$ delivery time due to finding the top-K items. Unfortunately, reducing a combinatorial search to a sort is still unsuitable for low-latency recommender systems with large catalogs. We present next the common solution to improve this. 

\textbf{(B) Two-stage recommendation:} Here we first generate a \emph{small} subset of potential item candidates $\mathcal{P}_{\rm sub} \subset \{1, \ldots, P\}$, and then select the top-K items in $\mathcal{P}_{\rm sub}$ leading to a $\mathcal{O}(|\mathcal{P}_{\rm sub}|)$ delivery time. 
This has two main shortcomings. First, the scoring model, which selects the highest scoring items from $\mathcal{P}_{\rm sub}$, does not directly optimize the reward for the whole slate, and rather optimizes a proxy offline metric for each item individually. This induces numerous biases related to the layout of the slate such as position biases where users tend to interact more often with specific positions \citep{biases_in_reco}. Second, the candidate generation and the scoring models are not necessarily trained jointly, which may lead to having candidates in $\mathcal{P}_{\rm sub}$ that are not the highest scoring items. 

\textbf{(C) Maximum inner product search (MIPS):} A practical approach to avoid the candidate generation step relies on approximate MIPS algorithms. Roughly speaking, these algorithms are capable of quickly sorting $P$ items in $\mathcal{O}(\log P)$ as long as the scores of items $a \in \{1, \ldots, P\}$ have the form $\bm{u}^\top \bm{\beta}_a$. Here $\bm{u} \in \real^d$ is a $d$-dimensional user embedding and $\bm{\beta}_a \in \real^d$ is the $d$-dimensional embedding of item $a$. This allows fast delivery of recommendation in roughly $\mathcal{O}(\log P)$ instead of $\mathcal{O}(P)$ without any additional candidate generation step. \textbf{PRR} uses approximate MIPS algorithms \citep{shrivastava2014asymmetric} making it suitable for extremely low-latency systems. We note that both IPS and DM can lead to a MIPS-compatible recommender system if the model (or the policy) is appropriately parametrized. However, much of the existing literature neglect this important consideration. In our case, decision making is reduced to a MIPS task and we use fast approximate algorithms to solve it. This allows us to avoid the two-stage scheme and its limitations. 

\paragraph{Summary of Limitations} Here we summarize the limitations of existing methods. \begin{enumerate}[label=\textbf{(\alph*)}]
    \item \textbf{Poor estimation of the reward:} this is due to the high variance and bias of IPS, the incorrect assumptions of existing DM and their potential modeling bias (e.g., only a single item is recommended, ignoring engagement features, etc.).
    \item \textbf{Policy learning for single items:} extending existing approaches to slates is complicated.
    \item \textbf{Slow decision making:} the real-time response requirement is not respected by most existing methods. The two-stage system is a remarkably pragmatic compromise. But it poses some challenges as we explained before. MIPS is a reliable practical alternative to two-stage systems but existing methods are usually not MIPS compatible.
    \end{enumerate}

\section{Algorithm} \label{sec:proposed_algorithm}
\begin{figure}
\centering
    \includegraphics[width=\linewidth]{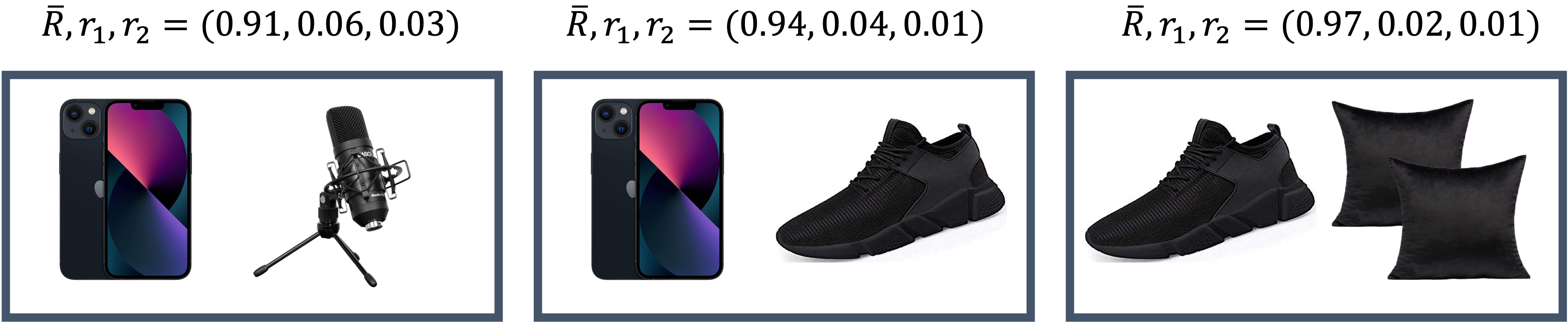}
    \caption{Example of 3 slates of size 2 on a technology website. From left to right are good, mixed and bad recommendations. $\bar{R}, r_1, r_2$ denote the probabilities of no-click, click on the $1$st and $2$nd  item, respectively.}
    \label{fig:examplebanners}
    \vspace{-0.6cm}
\end{figure}

\textbf{PRR} is a binary reward model that differentiates between item-level and slate-level features. The former reflects the quality of the slate as a whole while the latter is associated with the quality of individual items in the slate. This allows \textbf{PRR} to predict whether the user will interact with the slate (the reward) and which item will be interacted with (the rank). To see this, we give an example of the output of \textbf{PRR} in Figure \ref{fig:examplebanners}.  Here the user is interested in \textit{technology}. Then we show three slates of size 2.
In the left panel, the slate consists of two good\footnote{Here a good item refers to a technology item.} items: \textit{phone} and \textit{microphone}. The model predictions $(0.91, 0.06, 0.03)$ are the probabilities for no click, click on \textit{phone} and click on \textit{microphone}, respectively. The probability of a click on slate \textit{phone, microphone} is higher than the other slates and is equal to 0.09.  For comparison, the panel in the middle contains a good item  (\textit{phone}) in the prime first position but the \textit{shoe} in the second position, which is a poor match with the user interest in technology. As a consequence, the probabilities become $(0.94, 0.04, 0.01)$ for no click, click on \textit{phone} and click on \textit{shoe}. In the right panel, we show two poor items \textit{shoe} and \textit{pillow} resulting in the highest no-click probability $0.97$. 

The goal is to establish the level of association of each item (\textit{phone}, \textit{microphone}, \textit{shoe} and \textit{pillow}) with a particular user interest (\textit{technology}). At first glance, analyzing logs of successful and unsuccessful recommendations is the best possible way to learn this association. However, in practice, there are numerous factors that influence the probability of a click other than the quality of recommendations. In this example, the non-click probability of the good recommendations (\textit{phone}, \textit{microphone}) is 0.91 (click probability of 0.09), while the non-click probability of the bad recommendations (\textit{shoe}, \textit{pillow}) is 0.97 (click probability of 0.03). The change in the click probability from good to bad recommendations is relatively modest at only 0.06. Thus the model must capture additional factors that influence clicks. 

To account for this, \textbf{PRR} incorporates a real-world observation made by practitioners: typically the most informative features to predict successful interactions are \emph{engagement features}. These summarize how likely the user is to interact with the slate independently of the quality and relevance of its items to the user. This includes the slate size, its visibility and the level of user activity and engagement. While these features are strong predictors of interactions, they do not provide any information about which items are responsible for which interactions. In contrast, the \emph{recommendation features}, which include the user interests and the items shown in the slate, provide a relatively modest signal for predicting interactions. But they are very important for the recommendation task. Based on these observations, \textbf{PRR} leverages the engagement features to accurately learn the parameters associated with the \emph{useful} recommendation features. 

\textbf{PRR} also incorporates the information that different positions in the slate may have different properties. Some positions may boost a recommendation by making it more visible, and other positions may lessen the impact of the recommendation. To see this, consider the example in Figure \ref{fig:examplebanners}, the probability of clicking on \emph{shoes} increased by $0.01$ when placed in the prime first position (slate in the middle) compared to placing it in the second position (slate in the right). 


\subsection{Setting}\label{subsec:setting}
For any positive integer $P$, we define $[P]= \{1, 2, \ldots, P\}$. Vectors and matrices are denoted by bold letters. The $i$-th coordinate of a vector $\bm{x}$ is $x_i$; unless the vector is already indexed such as $\bm{x}_j$, in which case we write $x_{j, i}$. Let $\bm{A} \in \real^{P \times d}$ be a $P \times d$ matrix. Then for any $i \in [P]$, the $d$-dimensional vector corresponding to the $i$-th row of $\bm{A}$ is denoted by $\bm{A}_i \in \real^d$. Items are referenced by integers so that $[P]$ denotes the catalog of $P$ items. We define a \emph{slate} of size $K$, $\bm{s}= (s_\ell)_{\ell \in [K]}= \left(s_1, \ldots, s_K\right)$, as a $K$-permutation of $[P]$, which is an ordered collection of $K$ items from $[P]$. The space of all slates of size $K$ is denoted by $\mathcal{S}$. 

We consider a \emph{contextual bandit} setting where the agent interacts with users as follows. The agent observes a $d_x$-dimensional \emph{context} vector $\bm{x} \in \mathcal{X} \subseteq \real^{d_x}$. After that, the agent recommends a slate $\bm{s} \in \mathcal{S}$, and then receives a binary reward $R \in \{0, 1\}$ and a list of $K$ binary ranks $[r_1, \ldots, r_K] \in \{0, 1\}^K$ that depend on both the context $\bm{x}$ and the slate $\bm{s}$. The reward $R$ indicates whether the user interacted with the slate $\bm{s}$ and for any $\ell \in [K]$ the rank $r_\ell$ indicates whether the user interacted with the $\ell$-th item in the slate, $s_\ell$. The user can interact with \emph{at most} one item in the slate, and thus $R = \sum_{\ell \in [K]} r_\ell$. We let $\bar{R}=1-R$ so that $\bar{R} +  \sum_{\ell \in [K]} r_\ell=1$. Then the vector $(\bar{R}, r_1, \ldots, r_K) \in \real^{K+1}$ has one non-zero entry which is equal to one. 

We assume that the context $\bm{x}$ decomposes into two vectors as $\bm{x} = (\bm{y}, \bm{z})$ where $\bm{y} \in \real^{d^\prime}$ and $\bm{z} \in \real^{d_z}$. Here $\bm{y}$ denotes the engagement features that are useful for predicting the reward of a slate, independently of its items and the user interests. On the other hand, $\bm{z} \in \mathbb{R}^{d_z}$ denotes the remaining features in the context $\bm{x}$, which summarize the user interests. The dimensions of $\bm{z}$ and $\bm{x}$ are varying as they can contain the list of previously viewed items whose length may differ from one user to another. For this reason, these dimensions are subscripted by $\bm{z}$ and $\bm{x}$, respectively. In contrast, to simplify the notation, the dimension of $\bm{y}$, $d^\prime$, is fixed (although it can also be varying). We give a summary of our notation in Table \ref{tab:notation}. 

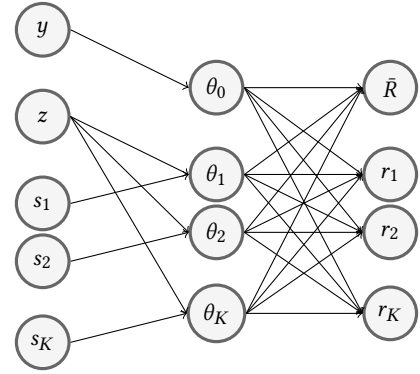
\begin{figure}
\begin{tikzpicture}[scale=0.77,
roundnode/.style={circle, draw=black!60, fill=black!4, very thick, minimum size=7mm}]
\node[roundnode] (inputy) at (0,2) {$y$};
\node[roundnode] (inputz) at (0,0.5) {$z$};
\node[roundnode] (inputs1) at (0,-1) {$s_1$};
\node[roundnode] (inputs2) at (0,-2) {$s_2$};
\node[roundnode] (inputsl) at (0,-3.4) {$s_K$};

\node[roundnode] (aaa) at (3,1) {$\theta_0$};
\node[roundnode] (bbb) at (3,-0.5) {$\theta_1$};
\node[roundnode] (ccc) at (3,-1.5) {$\theta_2$};
\node[roundnode] (ddd) at (3,-2.9) {$\theta_K$};

\node[roundnode] (Rbar) at (6,1) {$\bar{R}$};
\node[roundnode] (r1) at (6,-0.5) {$r_1$};
\node[roundnode] (r2) at (6,-1.5) {$r_2$};
\node[roundnode] (rk) at (6,-2.9) {$r_K$};

\draw[->] (inputy.east) -- (aaa.west);

\draw[->] (inputz.east) -- (bbb.west);
\draw[->] (inputs1.east) -- (bbb.west);

\draw[->] (inputz.east) -- (ccc.west);
\draw[->] (inputs2.east) -- (ccc.west);

\draw[->] (inputz.east) -- (ddd.west);
\draw[->] (inputsl.east) -- (ddd.west);

\draw[->] (aaa.east) -- (Rbar.west);
\draw[->] (aaa.east) -- (r1.west);
\draw[->] (aaa.east) -- (r2.west);
\draw[->] (aaa.east) -- (rk.west);

\draw[->] (bbb.east) -- (Rbar.west);
\draw[->] (bbb.east) -- (r1.west);
\draw[->] (bbb.east) -- (r2.west);
\draw[->] (bbb.east) -- (rk.west);

\draw[->] (ccc.east) -- (Rbar.west);
\draw[->] (ccc.east) -- (r1.west);
\draw[->] (ccc.east) -- (r2.west);
\draw[->] (ccc.east) -- (rk.west);

\draw[->] (ddd.east) -- (Rbar.west);
\draw[->] (ddd.east) -- (r1.west);
\draw[->] (ddd.east) -- (r2.west);
\draw[->] (ddd.east) -- (rk.west);

\end{tikzpicture}
\caption{A diagram of the PRR model.}
\label{prrdiagram}
\end{figure}

\subsection{Modeling Rank and Reward}\label{sec:modeling_reward_rank}
As we highlighted before, engagement features can be strong predictors of the reward of a slate independently of the quality of its items. Thus a model using these features while discarding the user interests might enjoy a high likelihood. But such a model is useless for personalized recommendation as it does not learn the user interests. This observation is often used to justify abandoning likelihood-based approaches in favor of ranking. Instead, \textbf{PRR} solves this issue by carefully incorporating both the engagement features $\bm{y}$, the user interests features $\bm{z}$ and the whole slate $\bm{s}$ to predict interactions accurately. The vector $(\bar{R}, r_1, \ldots, r_K) \in \real^{K+1}$ has exactly one non-zero entry which is equal to one. Thus we model it using a categorical distribution. Precisely, the \textbf{PRR} model has the following form 
\begin{align}\label{eq:prr_model}
\bar{R}, r_1, \ldots ,r_K | \bm{x}, \bm{s}  \sim {\rm cat}\left( \frac{\theta_0}{Z}, \frac{\theta_1}{Z}, \ldots, \frac{\theta_K}{Z} \right)\,,
\end{align}
where $Z = \sum_{k=0}^K \theta_\ell$, $\rm cat()$ is the categorical distribution, $\theta_0$ is the score of no interaction with the slate and $\theta_\ell$ is the score of interaction with the $\ell$-th item in the slate, $s_\ell$. The engagement features $\bm{y}$ are used to produce the positive score $\theta_0$ which is high if the chance of no interaction with the slate is high, independently of its items. It is defined as 
\begin{equation}\label{eq:theta0}
\theta_0 = \exp(\bm{y}^\top \bphi),
\end{equation}
where $\bphi$ is a vector of learnable parameters of dimension $d^\prime>0$. Similarly, let $\ell \in [K]$, the positive score $\theta_\ell$ is associated with the item in position $\ell$ in the slate, $s_\ell$, and is calculated in a way that captures user interests, position biases, and interactions that occur by \emph{accident}. Precisely, given a slate $\bm{s} = (s_\ell)_{\ell \in [K]} =  (s_1, \ldots, s_K)$ and user interests features $\bm{z}$, the score $\theta_\ell$ has the following form 
\begin{equation}\label{eq:thetak}
    \theta_\ell = \exp \{ g_{\bm{\Gamma}}(\bm{z})^\top \bPsi_{s_\ell}\} \exp(\gamma_{\ell}) + \exp(\alpha_{\ell}).
\end{equation}

Again this formulation is motivated by practitioners experience. The quantity $\exp(\alpha_\ell)$ denotes the additive bias for position $\ell \in [K]$ in the slate. It is high if there is a high chance of interaction with the $\ell$-th item in the slate irrespective of how appealing it is to the user. This quantity also explains interactions that are not associated at all with the recommendation (e.g., clicks by accident). The quantity $\exp(\gamma_\ell)$ is the multiplicative bias for position $\ell \in [K]$. It is high if a recommendation is \emph{boosted} by being in position $\ell \in [K]$. To see this, consider the example of ad placement and assume that we recommend a large slate of the form (\emph{phone,..., microphone}). Here \emph{phone} is placed in the first position while \emph{microphone} is placed in the last one. Now suppose that the user clicked on \emph{phone}. Then from a ranking perspective, we would assume that the user prefers the \emph{phone} over the \emph{microphone}. However, the user might have clicked on the \emph{phone} only because it was placed in the top position. \textbf{PRR} captures this through the multiplicative terms $\exp(\gamma_\ell)$.

The main quantity of interest is the recommendation score $g_{\bm{\Gamma}}(\bm{z})^\top \bPsi_{a}$ for $a \in [P]$. Here the vector $\bm{z} \in \real^{d_z}$ represents the user interests and the parameter vector $\bPsi_{s_\ell} \in \mathbb{R}^{d}$ represents the embedding of the $\ell$-th item in the slate, $s_\ell$. The vector $\bm{z}$ is first mapped into a fixed size $d$-dimensional embedding space using $g_{\bm{\Gamma}}(\cdot)$. The resulting inner product $g_{\bm{\Gamma}}(\bm{z})^\top \bPsi_{s_\ell}$ produces a positive or negative score that quantifies how good $s_\ell$ is to the user with interests $\bm{z}$. In practice, $\bm{z}$ can be the sequence of previously viewed items.

The \textbf{PRR} has multiple parameters $\bphi, \bm{\bm{\Gamma}}, \bPsi, \gamma_\ell,$ and $\alpha_\ell$ for $\ell \in [K]$. These are learned using the maximum likelihood principle where we assume access to logged data $\mathcal{D}_n$ of the form $\mathcal{D}_n=\{\bm{x}_i, \bm{s}_{i}, \bar{R}_i, r_{i, 1}, \ldots, r_{i, K} \, ; i \in [n]\}\,,$ such that $\bm{x}_i = (\bm{y}_i, \bm{z}_i)$ for any $i \in [n]$. With a slight abuse of notation, we will refer to the learned parameters by $\bphi, \bm{\Gamma}, \bPsi, \bm{\gamma}, \bm{\alpha}$ in the sequel. 
A neural network diagram of the \textbf{PRR} mode is shown in Figure~\ref{prrdiagram}.  It is enlightening to compare \textbf{PRR} with a fully connected network, unlike a fully connected network, only $y$ influences $\theta_0$, and for $l>0$, only $z$ and $s_l$ influence $\theta_l$.  These restrictions have two positive impacts 1) it enables recommendation by fast maximum inner product search, and 2) it reduces variance as the recommendation task is reduced to estimating an affinity between the user interests ($z$) and each recommendable item.  It also has one negative impact, the assumptions that there are no virtuous or detrimental combinations of recommendations is made increasing bias.

\subsection{Decision Making} 
From Equation \ref{eq:prr_model}, the probability of interaction with the slate is $P(R=1 \mid \bm{x}, \bm{s}) = 1- P(\bar{R}=1 \mid \bm{x}, \bm{s}) = 1-\frac{\theta_0}{Z}\,,
    = 1 - \frac{\theta_0}{\theta_0 +  \sum_{\ell \in [K]} \theta_\ell}$.
Then, from Equation \ref{eq:theta0}, \ref{eq:thetak}, the decision rule follows as
\begin{align}
   &{\rm argmax}_{\bm{s} \in \mathcal{S}} P(R=1 \mid \bm{x}, \bm{s})\nonumber\\
   &= {\rm argmin}_{\bm{s} \in \mathcal{S}} \frac{\theta_0}{\theta_0 +  \sum_{\ell \in [K]} \theta_\ell} \stackrel{(i)}{=} {\rm argmax}_{\bm{s} \in \mathcal{S}}  \sum_{\ell \in [K]} \theta_\ell\,,\nonumber\\
   &= {\rm argmax}_{\bm{s} \in \mathcal{S}}  \hspace{-0.1cm}\sum_{\ell \in [K]} \exp \{ g_{\bm{\Gamma}}(\bm{z})^\top \bPsi_{s_\ell}\} \exp(\gamma_{\ell}) + \exp(\alpha_{\ell})\,,\nonumber\\ &\stackrel{(ii)}{=} {\rm argmax}_{\bm{s} \in \mathcal{S}}  \sum_{\ell \in [K]} \exp \{ g_{\bm{\Gamma}}(\bm{z})^\top \bPsi_{s_\ell}\} \exp(\gamma_{\ell})\,, \label{eq:probability_click1}
\end{align}
where $(i)$ and $(ii)$ follow from the fact that both $\theta_0$ and $\exp(\alpha_{\ell})$ are additive and do not depend on $\bm{s}$. Our goal is to reduce decision making to a MIPS task. Thus the parametric form $\bm{u}^\top \bm{\beta}$ must be satisfied, which means that the sum $\sum_{\ell \in [K]}$, the exponential in $\exp \{ g_{\bm{\Gamma}}(\bm{z})^\top \bPsi_{s_\ell}\}$ and the term $\exp(\gamma_{\ell})$ in Equation \ref{eq:probability_click1} need to be removed. This is achieved by first sorting the position biases as
\begin{align}
   i_1, \ldots, i_K = {\rm argsort}(\gamma)\,. \label{eq:pos_biases_sort}
\end{align}
This is done once since Equation \ref{eq:pos_biases_sort} neither depend on the items nor the user. Then MIPS is performed as  
\begin{align}
   s_1^\prime, \ldots, s_K^\prime = {\rm argsort}(g_{\bm{\Gamma}}(\bm{z})^\top \bPsi)_{1:K}\,. \label{eq:mips1}
\end{align}
Finally, the recommended slate $\bm{s} = (s_1, s_2, \ldots, s_K)$ is obtained by rearranging the items $ s_1^\prime, \ldots, s_K^\prime$ as
\begin{align}
s_1, s_2, \ldots, s_K  = s_{i_1}^\prime, s_{i_2}^\prime, \ldots, s_{i_K}^\prime\,. \label{eq:recommended_slate}
\end{align}
In other terms, we select the top-K items with the highest recommendation scores $g_{\bm{\Gamma}}(\bm{z})^\top \bPsi_a$ for $a \in [P]$. We then place the highest scoring item into the best position, that is the position $\ell \in [K]$ with the largest value of $\gamma_\ell$. Then the second-highest scoring item is placed into the second-best position, and so on. The procedure in Equation \ref{eq:mips1} ,\ref{eq:pos_biases_sort},\ref{eq:recommended_slate} is equivalent to the decision rule in \cref{eq:probability_click1}. It is also much more computationally efficient as \cref{eq:mips1} can be performed roughly in $\mathcal{O}(\log P)$ thanks to fast approximate MIPS algorithms \citep{shrivastava2014asymmetric}, while Equation \ref{eq:probability_click1} requires roughly $\mathcal{O}(P^K)$. The time complexity is also improved compared to ranking approaches by $\mathcal{O}(P / \log P)$. This makes \textbf{PRR} scalable to huge action spaces.  

Note that $\bphi, \bm{\alpha}$ are \emph{nuisance} parameters as they are not needed for decision making; only the recommendation scores $g_{\bm{\Gamma}} (\bm{z})^\top \Psi_a$ and the multiplicative position biases $\exp(\gamma_\ell)$ are used in the procedure in Equation \ref{eq:mips1},\ref{eq:pos_biases_sort}, \ref{eq:recommended_slate}. While not used in decision making, learning these parameters is necessary to accurately predict the recommendation scores. Also, including them in the model provides room for interpretability in some cases.

To summarize, \textbf{PRR} has the following properties. \begin{enumerate*}[label=\textbf{\arabic*)}]
\item It models the joint distribution of the reward and ranks $(\bar{R}, r_1, \ldots, r_K)$ in the simple formulation in Equation \ref{eq:prr_model}.
\item It makes use of engagement features $\bm{y}$ in order to help learn the recommendation signal more accurately.
\item Its recommendation scores have a parametric form that is suitable for MIPS, which allows fast decision making in $\mathcal{O}(\log P)$.
\end{enumerate*}

When compared to prior works, \textbf{PRR} is uniquely designed for scenarios in which users examine the entire slate and can interact with at most one of its items. This realistic and important setting sets \textbf{PRR} apart from prior works. Additionally, \textbf{PRR} combines the strengths of both reward and ranking approaches. Reward-based methods \citep{dudik2014doubly} focus primarily on optimizing the reward signal, aligning offline optimization with A/B test outcomes. However, they overlook the rank signal, resulting in challenges for learning, particularly in large-scale tasks. Conversely, ranking approaches \citep{rendle2012bpr} rely on heuristics centered around proxy scores for individual items, which may not accurately reflect A/B test results \citep{Garcin2014}. \textbf{PRR} bridges these two paradigms by directly optimizing the reward while also leveraging the rank signal.

\section{Experiments} \label{sec:experiments}

We evaluate \textbf{PRR} using synthetic and real-world problems that mimic the sequential interactions between users and recommender systems. The other alternatives consist in either using information-retrieval metrics or IPS. Unfortunately, the former is not aligned with online A/B test results, while the latter can suffer high bias and variance in large-scale settings \citep{reco_simulation}. Next we briefly present our experimental design while we defer further detail to the Appendix.

\paragraph{Goals} Our paper is tailored to the common scenario where the user examines an entire slate and can interact with at most one of its items. Accordingly, our experiments aim to demonstrate the following. \textbf{1)} The effectiveness of our model in this specific setting. \textbf{2)} The benefit of combining both the reward and rank signals by comparing \textbf{PRR} to other variants  that solely employ one of these signals. \textbf{3)} The benefit of incorporating engagement features by comparing \textbf{PRR} to a variant that excludes them.

\subsection{Baselines}
 \textbf{PRR} optimizes the reward offline and thus we only compare it to off-policy reward optimizing approaches. This does not include collaborative filtering \citep{su2009survey}, content-based \citep{lops2011content}, or on-policy reinforcement learning methods \citep{ie2019slateq}. Thus we mainly compare \textbf{PRR} to the methods reviewed IPS and DM discussed in related work. As existing DMs are crafted for scenarios that are different from ours, detailed in our related work section (e.g., the user can click on multiple items in the slate), we focus our comparison on IPS and its variants. These methods are agnostic to the reward generation process, unlike DMs that assume specific reward structures which don't apply to our setting. However, we include three DMs derived from \textbf{PRR} which are used to validate some of our modeling assumptions. Finally, for completeness, we also include two widely used click models, cascading models (\textbf{CM}) and position-based models (\textbf{PBM}) \citep{kiyohara2022doubly,cief2022pessimistic}.
 
 \textbf{Variants of \textbf{PRR}:} we consider three variants of \textbf{PRR}. First, \textbf{{PRR-reward}} uses only the reward and ignores the rank. \textbf{{PRR-reward}} is trained on both, successful and unsuccessful slates. Second, \textbf{{PRR-rank}} only uses the rank and is consequently trained on successful slates only. Finally, \textbf{{PRR-bias}} ignores the engagement features $\bm{y}$ and sets $\theta_0 = \exp(\phi)$ where $\phi$ is a scalar parameters ($\phi$ replaces $\bm{y}^\top \bphi$). Comparing \textbf{PRR} to \textbf{{PRR-reward}} and \textbf{{PRR-rank}} is to show the benefits of combining both signals, while comparing it to \textbf{{PRR-bias}} it to highlight the effect and importance of the engagement features $\bm{y}$. The three models are summarized below.
\begin{align*}
& \textbf{{PRR-reward:}} \\  &\bar{R}, r_1, \ldots ,r_K | \bm{x}, \bm{s}  \sim {\rm cat}\Big( \frac{\theta_0}{Z}, \frac{\sum_{\ell=1}^K\theta_\ell}{Z} \Big), \quad Z = \sum_{\ell=0}^K \theta_\ell\,,\\
&\textbf{{PRR-rank:}} \\ & r_1,\ldots,r_K \mid \bm{x}, \bm{s}  | {\rm cat}\Big( \frac{\theta_1}{Z}, \ldots, \frac{\theta_K}{Z} \Big), \quad Z = \sum_{\ell=1}^K \theta_\ell\,,\\
&\textbf{{PRR-bias:}} \\ & \bar{R}, r_1, \ldots ,r_K | \bm{x}, \bm{s}  \sim {\rm cat}\Big( \frac{\phi}{Z}, \frac{\theta_1}{Z}, \ldots, \frac{\theta_K}{Z} \Big), Z \hspace{-0.1cm}= \hspace{-0.1cm}\phi  + \hspace{-0.1cm}\sum_{\ell=1}^K \theta_\ell.
\end{align*}
where $\theta_0$ and $\theta_\ell$ for $\ell \in [K]$ are defined in Equation \ref{eq:theta0} - \ref{eq:thetak} while $\phi \in \real$ in \textbf{{PRR-bias}} is a learnable parameter.

\textbf{Inverse propensity scoring:} We also consider IPS estimators of the expected reward of policies that are designed by removing the preference bias of the logging policy $\pi_0$ in data $\mathcal{D}_n$. This is achieved by re-weighting samples using the discrepancy between the learning policy $\pi$ and the logging policy $\pi_0$ such as
\begin{align}\label{eq:ips_policy_value}
  \hat{V}_n^{\rm IPS}(\pi) &= \frac{1}{n} \sum_{i=1}^n R_i \frac{\pi(\bm{s}_i | \bm{z}_i)}{\pi_0(\bm{s}_i | \bm{z}_i)}\,.
\end{align}
This estimator is unbiased when $\pi$ and $\pi_0$ have common support. But it can be highly biased when this assumption is violated, which is common in practice. It also suffers high variance. One way to mitigate this is to reduce the action space from slates to items \citep{li2018offline}. This is achieved by assuming that the reward $R$ is the sum of rank $r_1, \ldots, r_K$, and that the $\ell$-th rank, $r_\ell$, only depends on the item $s_\ell$ and its position $\ell$. This allows estimating the expected reward of the learning policy $\pi$ as
\begin{align}\label{eq:iips_policy_value}
    \hat{V}^{\rm IIPS}(\pi) &= \frac{1}{n} \sum_{i=1}^n \sum_{\ell=1}^K r_{i,\ell} \frac{\pi(s_{i,\ell}, \ell | \bm{z}_i)}{\pi_0(s_{i,\ell}, \ell | \bm{z}_i)}\,,
\end{align}
where $\pi(a, \ell | \bm{z})$ and $\pi_0(a, \ell | \bm{z})$ are the marginal probabilities that the learning policy $\pi$ and the logging policy $\pi_0$ place the item $a$ in position $\ell \in [K]$ given user interests $\bm{z}$, respectively. Note that in practice computing these marginals is often intractable for both $\pi$ and $\pi_0$; in which case approximation must be employed.

The next step is to optimize the estimator ($\hat{V}_n^{\rm IPS}(\pi)$ or $\hat{V}_n^{\rm IIPS}(\pi)$) to find the policy that will be used for decision making. To achieve this, we need to parameterize the learning policy $\pi$. Here we assume that $\pi$ is parametrized as a factored softmax
\begin{align}\label{ips_parametrization}
&\pi(\bm{s} \mid \bm{z}) = \pi_{\Xi,\beta,K}(\bm{s}|\bm{z}) = \prod_{\ell=1}^K p_{\Xi,\beta}(s_\ell|\bm{z})\,,\\
&\text{where } \, p_{\Xi,\beta}(a|\bm{z}) = \frac{\exp \{ f_{\bm{\Xi}}(\bm{z})^\top \bm{\beta}_a\} }{\sum_{a^\prime \in [P]} \exp \{ f_{\bm{\Xi}}(\bm{z})^\top \bm{\beta}_{a^\prime} \}}\,,\nonumber
\end{align}
where $\bm{\beta}_a \in \real^d$ is the embedding of item $a$ and $f_{\bm{\Xi}}$ maps user interests $\bm{z} \in \real^{d_z}$ into a $d$-dimensional embedding. Finally, $\hat{V}_n^{\rm IIPS}(\pi)$ also requires the marginal probabilities $\pi(a, \ell | \bm{z})$ and $\pi_0(a, \ell | \bm{z})$. In our case, $\pi(a, \ell | \bm{z}) = p_{\Xi,\beta}(a|\bm{z})$ while we may need to approximate $\pi_0(a, \ell | \bm{z})$ depending on the logging policy. While convenient, factored policies have significant limitations. In particular, IIPS with factored policies might cause the learned policy to converge to selecting slates with repeated items, which is illegal. Thus to be fair to IIPS, we use sampling \emph{without replacement} in decision making. Another alternative to mitigate this is 
the top-K heuristic \citep{chen2019top} which causes the probability mass in $\pi_{\Xi,\beta,K}(\bm{s}|\bm{z})$ to be spread out over the top-K high scoring items rather than a single one. We denote the IIPS estimator combined with the top-K heuristic by \textbf{top-K IIPS}.

\subsection{Synthetic Problems}\label{synthetic_data}

We design a simulated A/B test protocol that takes different recommender systems as input and outputs their respective reward. We first define the problem instance consisting of the true parameters (oracle) and the logging policy as $\{ \bphi, \bPsi, \bm{\gamma}, \bm{\alpha}, g_{\bm{\Gamma}}(\cdot), P_{\bm{y}}(\cdot), P_{\bm{z}}(\cdot), P_{K}(\cdot)\}$ and $\pi_0$. Here $P_{\bm{y}}(\cdot), P_{\bm{z}}(\cdot)$, and $P_{K}(\cdot)$ are the distributions of the engagement features, the user interests features and the slate size, respectively. Given the oracle, we produce offline training logs and propensity scores $\{ \mathcal{D}, \mathcal{P} \}$ by running the logging policy $\pi_0$ in our simulated environment and observing its reward and rank. These logs are then used to train \textbf{PRR} and the baselines. After training, a simulated A/B test is used for testing. We defer a detailed description of our simulation environment for reproducibility to the Appendix. For instance, we summarize in Figure \ref{alg:simulated_logs} the data generation process while we present in Figure \ref{alg:abtest} the simulated A/B test.

We consider two non-personalized logging policies. \begin{enumerate*}[label=\textbf{(\alph*)}]
    \item \textbf{{uniform:}} this policy samples uniformly without replacement $K$ items from the catalog $[P]$. That is $\pi_0(\bm{s} \mid \bm{z}) = \frac{1}{P (P-1)\ldots (P - K+1)}$ for any slate $\bm{s} \in \mathcal{S}$ and any user interests $\bm{z}$. The marginal distribution can be computed in closed-form as $\pi_0(a, \ell | \bm{z}) = 1/P$.
    \item \textbf{{top-K pop:}} this policy samples without replacement $K$ items where the probability of an item $a$ is proportional to the $L_2$ norm of its embedding, $\|\bm{\Psi}_a\|$. This is based on the intuition that a large value of $\|\bm{\Psi}_a\|$ means that item $a$ is recommended more often and thus it is more popular. We stress that this logging policy has access to the true embeddings $\bm{\Psi}$ of the simulated environment (Figure \ref{alg:simulated_logs}). Here the marginal distribution is intractable and we simply approximate is as $\pi_0(a, \ell | \bm{z}) \approx \|\Psi_a\|/{\sum_{a^\prime} \|\Psi_{a^\prime}\|}$ for IIPS.
     \end{enumerate*}

\begin{figure}
\includegraphics[width=\linewidth]{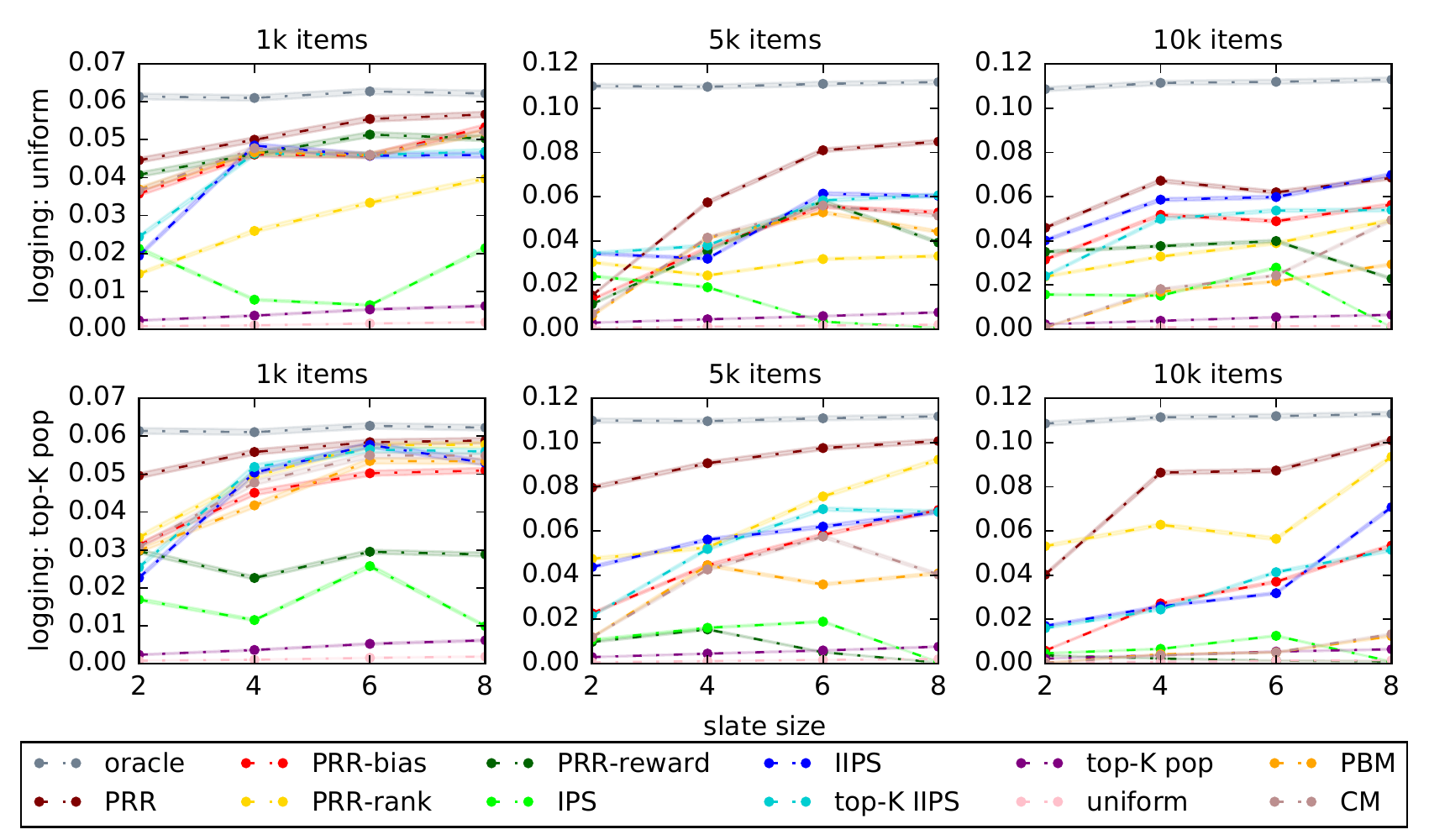}
\caption{The reward (y-axis) of methods in \textbf{synthetic problems} with varying slate sizes (x-axis), number of items (columns) and logging policies (rows). The shaded areas represent uncertainty and they are small since we run long A/B tests with $n_{\rm test}=100{\rm k}$.}
\label{fig:simulated_ab_test}
\end{figure}

In Figure \ref{fig:simulated_ab_test}, we report the average A/B test reward of \textbf{PRR} with varying slate sizes, number of items and logging policies using 100k training samples. Overall, we observe that \textbf{PRR} outperforms the baselines across the different settings. Next we summarize the general trends of algorithms. 
\begin{enumerate}[label=\textbf{(\alph*)},topsep=0pt,itemsep=0pt,partopsep=0pt,parsep=1ex]
\item \textbf{Varying logging policy:} models that use the reward only, \textbf{{IPS}} and \textbf{{PRR-reward}}, favor uniform logging policies while those that use only the rank, \textbf{{IIPS}} and \textbf{{PRR-rank}} perform better with the \textbf{top-K pop} logging policy. \textbf{{PRR-bias}} discards the slate-level features $\bm{y}$ and uses a single parameter $\phi$ for all slates. Thus \textbf{{PRR-bias}} benefits from uniform logging policies as they allow learning $\phi$ that works well across all slates. Indeed, in Figure \ref{fig:simulated_ab_test} the gap between \textbf{PRR} and \textbf{{PRR-bias}} shrinks for the uniform logging policy. Finally, the performance of \textbf{PRR} is relatively stable for both logging policies.
\item \textbf{Varying slate size:} the performance of models that use the reward only, \textbf{{IPS}} and \textbf{{PRR-reward}}, deteriorates when the maximum slate size increases. On the other hand, those that use only the rank, \textbf{{IIPS}} and \textbf{{PRR-rank}}, benefit from larger slates as this leads to displaying more item comparisons. The addition of the \textbf{{top-K}} heuristic improves the performance of \textbf{{IIPS}} in some cases by spreading the mass over different items, making it not only focus on retrieving one but several high scoring items. However, the increase in performance is not always guaranteed which might be due to our choice of hyperparameters or our approximation of the marginal distributions of policies. Finally, \textbf{PRR} performs well across all slate sizes as it uses both the reward and rank.
\item \textbf{Varying number of items:} the models that use the rank benefit from large slates. Here we observe that the increase in performance is more significant for large catalogs. In contrast, models that use only the reward suffer a drop in performance when the number of items increases.
\end{enumerate}

\subsection{Session Completion Problems} \label{semi_synthetic_data}
We use the Twitch dataset \citep{rappaz2021recommendation} to evaluate \textbf{PRR} on user session completion tasks. Roughly speaking, we process the dataset such that each user $u$ has a list $\mathcal{I}_u$ that contains the items that the user interacted with. We randomly split these user-item interactions $\mathcal{I}_u$ into two parts, $\mathcal{I}_u^{\textsc{view}}$ and $\mathcal{I}_u^{\textsc{hide}}$. $\mathcal{I}_u^{\textsc{view}}$ is observed while $\mathcal{I}_u^{\textsc{hide}}$ is hidden. The baselines are evaluated based on their ability to predict the hidden session of a user $\mathcal{I}_u^{\textsc{hide}}$ by only observing a part of it, $\mathcal{I}_u^{\textsc{view}}$.


Logged data $\mathcal{D}_n$ is collected using the \textbf{{top-K pop}} logging policy. In each iteration $i \in [n]$, we randomly sample a user $u_i$. Then, we recommend a slate $\bm{s}_i = (s_{i, \ell})_{\ell \in [K]}$ by sampling without replacement $K$ items with probabilities proportional to their popularity, i.e., their number of occurrences in the dataset. After recommending $\bm{s}_i = (s_{i, \ell})_{\ell \in [K]}$, we construct a binary vector $b_i = (\mathbb{I}_{\{s_{i, \ell} \in \mathcal{I}_u^{\textsc{hide}}\}})_{\ell \in [K]}  = (\mathbb{I}_{\{s_{i, 1} \in \mathcal{I}_u^{\textsc{hide}}\}}, \mathbb{I}_{\{s_{i, 2} \in \mathcal{I}_u^{\textsc{hide}}\}}, \ldots,\mathbb{I}_{\{s_{i, K} \in \mathcal{I}_u^{\textsc{hide}}\}}) \in \real^K$. In other words, for any $\ell \in [K]$, $b_{i,\ell}=1$ if the $\ell$-th item in the slate, $s_{i,\ell}$, is in the hidden user-item interaction $\mathcal{I}_u^{\textsc{hide}}$, and $b_{i,\ell}=0$ otherwise. This binary vector $b_i$ is then used to generate the reward and rank signals $R_i, r_{i, 1}, \ldots, r_{i, K}$ for user $u_i$ and slate $\bm{s}_i$. This allows constructing logged data $\mathcal{D}_n$. After training the baselines on $\mathcal{D}_n$, they are evaluated following the data collection process except that they make decisions instead of the logging policy $\pi_0$. More details about the data generation and testing processes are given in the Appendix.

In Figure \ref{fig:semi_synthetic} (left-hand side), we report the results of \textbf{PRR} and the baselines on user session completion tasks. Note that there are no engagement features in this problem. Thus \textbf{PRR} is the same as \textbf{{PRR-bias}} and hence we only include \textbf{PRR} in \cref{fig:semi_synthetic}. Furthermore, in our synthetic problems, we used implementations of the cascading models (\textbf{CM}) and position-based models (\textbf{PBM}), assuming a binary context $\bm{z}$, which was appropriate for those instances. However, this session completion problem encompasses non-binary contexts, and thus these models have been omitted from \cref{fig:semi_synthetic}. It's important to highlight that \textbf{CM} and \textbf{PBM} are not the most competitive baselines in our synthetic problems and are not well-suited for our specific setting. Overall, \textbf{PRR} outperforms the other baselines in the session completion problem. We also observe that \textbf{{PRR-reward}} has good performance in this scenario, while all the other methods have comparable performance which is lower than that of \textbf{{PRR-reward}}.

\begin{figure}
\centering
\includegraphics[width=\linewidth]{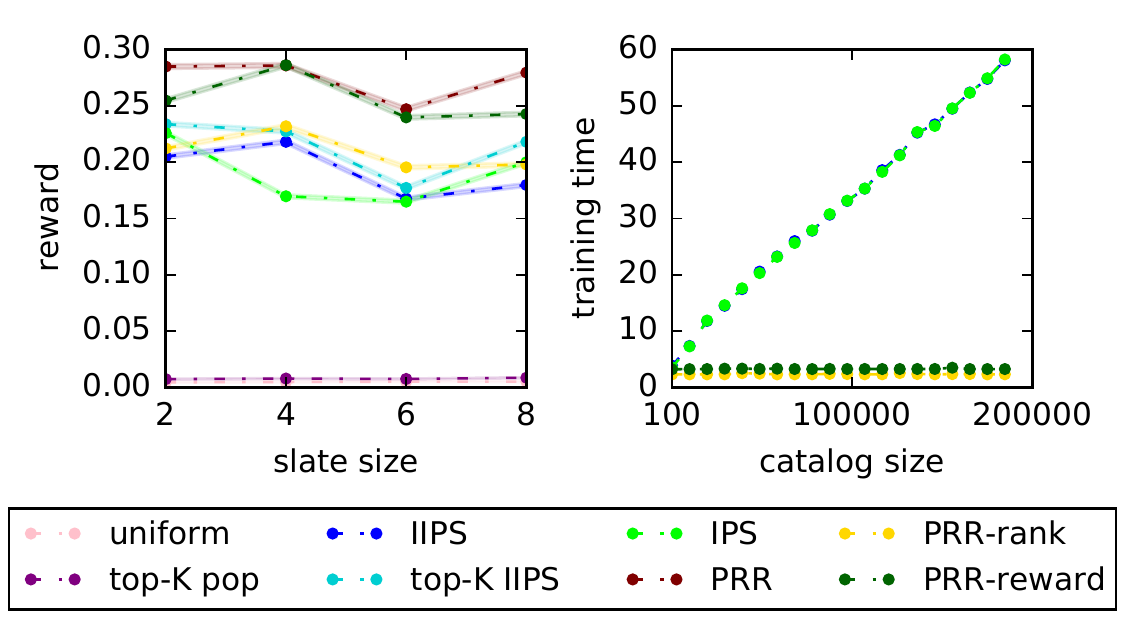}
\caption{\textbf{On the left-hand side}, we report the reward (y-axis) of methods in \textbf{session completion problems} with varying slate sizes (x-axis). \textbf{On the right-hand side}, we report the training time (y-axis) of \textbf{PRR} and baselines in \textbf{session completion problems} with varying catalog sizes (x-axis).}
\label{fig:semi_synthetic}
\end{figure}

\subsection{Computational Efficiency}

We assess the training speed of the algorithms with respect to the catalog and slate sizes $P$ and $K$. First, \textbf{PRR} and its variants compute $K+1$ scores $\theta_0, \dots, \theta_K$ and normalize them using $Z=\sum_{\ell=0}^K \theta_\ell$. Therefore, evaluating \textbf{PRR} and its variants in one data-point costs roughly $\mathcal{O}(K)$, where we omit the cost of computing the scores since it is the same for all algorithms. In contrast, \textbf{{IPS}} and its variants compute a softmax over the catalog. This requires computing the normalization constant $\sum_{a' \in [P]} \exp \{ f_{\bm{\Xi}}(\bm{z})^\top \bm{\beta}_{a'} \}$ in Equation \ref{ips_parametrization}. Thus the evaluation cost of \textbf{{IPS}} and its variants is roughly $\mathcal{O}(P)$. This is very costly compared to $\mathcal{O}(K)$ in realistic settings where $P \gg K$. An additional consideration to compare the training speed of algorithms is whether they use successful slates only, which significantly reduces the size of training data. Taking this into account, the fastest of all algorithms is the \textbf{{PRR-rank}} since its evaluation speed is $\mathcal{O}(K)$ and it is trained on successful slates only. In our experiments, \textbf{{PRR-rank}} is $\bm{\approx 20 \times}$ faster to train than \textbf{{IPS}}. These computational dependencies on $K$ and $P$ are also highlighted in Figure \ref{fig:semi_synthetic} (right-hand side). In particular, \textbf{{IPS}} training time scales linearly in $P$ while \textbf{{PRR}} do not have such a dependency on $P$.

\begin{table}
\centering
\caption{Properties of \textbf{PRR} and the baselines.}
\begin{tabular}{l c c}
  \toprule
  \textbf{Method} & \textbf{Computational} & \textbf{Empirical }\\ 
   & \textbf{efficiency} & \textbf{performance}\\   
  \bottomrule
    \textbf{PRR} & $\mathcal{O}(K)$ & High \\
    \textbf{{PRR-bias}} & $\mathcal{O}(K)$ & Medium \\
    \textbf{{PRR-rank}} & $\mathcal{O}(K)$ & Medium \\
    \textbf{{PRR-reward}} & $\mathcal{O}(K)$ & Low \\
    \textbf{{IPS}} & $\mathcal{O}(P)$ & Low\\
    \textbf{{IIPS}} & $\mathcal{O}(P)$ & Medium\\
    \textbf{{Top-K IIPS}} & $\mathcal{O}(P)$ & Medium\\
    \bottomrule
  \end{tabular}
  \label{comparison}
\end{table}

\subsection{Limitations of \textbf{PRR}}\label{prr_limitations}
After presenting and evaluating our algorithm, we are in a position to discuss its limitations. \begin{enumerate}[label=\textbf{(\alph*)}]
    \item \textbf{Modeling capacity:} \textbf{PRR} is trained to predict the reward of any slate which is a complex task. Thus high-dimensional embeddings might be needed to make accurate and calibrated predictions. As a result, the MIPS task produced by \textbf{PRR} might require high dimensional embeddings that do not conform to engineering constraints. A possible path to mitigate this is to observe that accurately predicting the reward is sufficient but not necessary for the recommendation task. When recommending, we only need to find the best slate. This is a simpler task and might be achieved with much smaller embeddings than those required by \textbf{PRR}. Therefore, one can train \textbf{PRR} with high-dimensional embeddings. Then optimize policies parametrized with low-dimensional embeddings using the learned reward estimates of \textbf{PRR} instead of IPS. The learned policy that fits the engineering constraints will be then used in decision making.
    \item \textbf{Incorporating prior information:} we learned the parameters of \textbf{PRR} using the maximum likelihood principle. While powerful, this does not allow incorporating prior information about actions and users into our model. Therefore, a natural extension of \textbf{PRR} is to use a Bayesian approach by adding a prior distribution that captures similarities between actions \citep{aouali2023mixed}. 
    \item \textbf{Theoretical analysis:} one of the main appeals of IPS is that it can be analyzed theoretically. This is difficult for \textbf{PRR} although a Bayesian analysis where we assume that the model of the environment is the same as that of \textbf{PRR} might be possible \citep{aouali2023mixed}. 
    \end{enumerate}

\section{Conclusion} \label{sec:conclusion}

We present \textbf{PRR}, a scalable probabilistic model for personalized slate recommendation. \textbf{PRR} efficiently estimates the probability of a slate being successful by combining the reward and rank signals. It also optimizes the reward of the whole slate by distinguishing between slate-level and item-level features. Experiments attest that \textbf{PRR} outperforms the baselines and it is far more scalable, in both training and decision making. 

As a parametric model \textbf{PRR} is based upon assumptions, in particular, the user engagement, user interest and recommended items are restricted in their ability to influence the no click or click on a particular item probability (see Figure~\ref{prrdiagram}).  Lifting these assumptions would be a promising avenue for research, and this would open the door to models that allow virtuous or detrimental combinations of recommendations.  Any model that considers these interactions must identify novel ways to a) deal with the massive increase in variance from measuring the combinatorial explosion in possible interaction effects, and b) solve a combinatorial optimization problem at recommendation time.  We leave these important questions to future work.

\bibliographystyle{ACM-Reference-Format}
\bibliography{main}

\newpage

\appendix

\section{Summary of Notation}\label{app:notation}

We provide a summary of our notation in \cref{tab:notation}.

\begin{table}[H]
\centering
\caption{Notation.}
\begin{adjustbox}{max width=\linewidth}
\begin{tabular}{l l}
  \toprule
  \textbf{Notation} & \textbf{Definition} \\ 
  \bottomrule
  $\bm{x} = (\bm{y}, \bm{z}) \in \mathbb{R}^{d_x}$ & context. \\
$\bm{y} \in \mathbb{R}^{d^\prime}$ & engagement features. \\
$\bm{z} \in \mathbb{R}^{d_z}$ & user interests features. \\
$R \in \{0, 1\}$ & reward indicator.\\
$\bar{R} \in \{0, 1\}$ & regret indicator.\\
$r_\ell \in \{0, 1\}$ & rank indicator of the item in position $\ell \in [K]$.\\
$\bphi \in \mathbb{R}^{d^\prime}$ & engagement parameters.\\
$\gamma_\ell \in \real$  & multiplicative position bias in position  $\ell \in [K]$.\\
$\alpha_\ell \in \real$  & additive position bias in position $\ell \in [K]$.\\
$g_{\bm{\Gamma}}(\bm{z}) \in \real^d$ & user embedding.\\
$\bPsi \in \real^{P \times d}$ & items embeddings.\\
$\theta_0 \in \real$ & score for no-interaction with the slate.\\ 
$\theta_\ell \in \real$ & score for an interaction with the item in position $\ell \in [K]$. \\ 
$\bm{s} = (s_1,...,s_K)$ & slate of $K$ recommendations $s_\ell \in [P]$ for any $\ell \in [K]$.\\
    \bottomrule
  \end{tabular}
  \end{adjustbox}
\label{tab:notation}
\end{table}

\section{Experimental Design}\label{app:experimental_design}
Here we give more information about our experiments. We start with the synthetic problems and then present the session completion problems.

\subsection{Synthetic Problems}\label{app:synthetic}

We design a simulated A/B test protocol that takes different recommender systems as input and outputs their respective reward. We first define the problem instance consisting of the true parameters (oracle) and the logging policy as $\{ \bphi, \bPsi, \bm{\gamma}, \bm{\alpha}, g_{\bm{\Gamma}}(\cdot), P_{\bm{y}}(\cdot), P_{\bm{z}}(\cdot), P_{K}(\cdot)\}$ and $\pi_0$. Here $P_{\bm{y}}(\cdot), P_{\bm{z}}(\cdot)$ and $P_{K}(\cdot)$ are the distributions of the engagement features, the user interests features and the slate size, respectively. Given the oracle, we produce offline training logs and propensity scores $\{ \mathcal{D}, \mathcal{P} \}$ by running the logging policy $\pi_0$ as described in \cref{alg:simulated_logs}. These logs are then used to train \textbf{PRR} and competing baselines. After training, the simulated A/B test in \cref{alg:abtest} is used for testing. 

In all our experiments, the true parameters are sampled randomly as 
\begin{align*}
   & \bphi \sim \mathcal{N}(\mu_\phi, \Sigma_\phi) \,, \quad \bPsi \sim \mathcal{N}(\mu_\psi, \Sigma_\psi)\,, \quad \bm{\Gamma} \sim \mathcal{N}(\mu_\Gamma, \Sigma_\Gamma)\,,\\  &\bm{\gamma} \sim \mathcal{N}(\mu_\gamma, \Sigma_\gamma) \,, \quad \bm{\alpha} \sim \mathcal{N}(\mu_\alpha, \Sigma_\alpha)\,.
\end{align*}
For each user, the engagement features $\bm{y}$ are sampled randomly following the distribution $P_{\bm{y}} = \mathcal{N}(\mu_y, \Sigma_y)$. For the interest features $\bm{z}$, we assume that there are $L$ topics and $\bm{z}$ is consequently generated as follows. For each user $u$, we randomly sample the number of topics that interest user $u$ as $L_u \sim 1 + \mathcal{P}oison(3)$. After that, we uniformly sample $L_u$ topics that interest the user from $[L]$. It follows that $\bm{z} \in \real^L$ ($d_z=L$) is represented as a binary vector such as $z_\ell=1$ if the user is interested in topic $\ell$, and $z_\ell=0$ otherwise. For simplicity, the mapping $g_{\bm{\Gamma}}$ is linear and defined as $g_{\bm{\Gamma}}(\bm{z}) = \bm{\Gamma}\bm{z}$. Finally, for each user, the slate size is sampled uniformly from $[K]$ where $K$ is the maximum slate size. For reproducibility, the \texttt{SEED} is fixed at $42$ which was chosen randomly. We considered realistic settings with relatively large catalogs and slates. The catalog size varies between $1000$ and $10000$ and the slate size varies between $2$ and $8$. \textbf{PRR} is suitable for larger catalogs but the baselines become very slow to train. This explains our choice of $10000$ as a maximum catalog size. In optimization, we use Adam \citep{kingma2014adam} with a learning rate of 0.005 for 100
epochs using mini-batches of size 516.

\SetKwInput{KwInput}{Input}
\SetKwInput{KwOutput}{Output}

\RestyleAlgo{ruled}
\begin{algorithm}
    \caption{Synthetic training logs}
    \label{alg:simulated_logs}
    \KwInput{oracle parameters $\{ \bphi, \bPsi, \bm{\gamma}, \bm{\alpha}, g_{\bm{\Gamma}}(\cdot), P_{\bm{y}}(\cdot), P_{\bm{z}}(\cdot), P_{K}(\cdot)\}$, logging policy $\pi_0(\bm{s} \mid \bm{x})$, marginal logging policies $\pi_0(s_1|\bm{x}),\ldots,\pi_0(s_K|\bm{x})$, number of training samples $n_{\rm train}$.}
    \KwOutput{logs $\mathcal{D}$, propensity scores $\mathcal{P}$.}
    $\mathcal{D} \gets \{ \,\}\,, \quad \mathcal{P} \gets \{ \,\}$\\
     \For{$i=1, \ldots, n_{\rm train}$}{
        $\bm{y}_i \sim P_{\bm{y}}(\cdot)\,, \quad \bm{z}_i \sim P_{\bm{z}}(\cdot)\,, \quad K_i \sim P_K(\cdot)$\\
        $ \bm{s}_i = (s_{i,1},\ldots, s_{i,K_i}) \sim \pi_0( \cdot |\bm{z}_i)$\\
        $\theta_0 \gets \exp(\bm{y}_i^\top \bphi)$\\
        \For{$\ell=1, \ldots, K_i$}{$\theta_\ell \gets \exp (g_{\bm{\Gamma}}(\bm{z}_i)^\top \bPsi_{s_{i, \ell}}) \exp(\gamma_\ell) + \exp(\alpha_\ell)$}
        $\bar{R}_{i}, r_{i,1},\ldots, r_{i,K}  \sim {\rm cat}\left(\frac{\theta_0}{Z}, \frac{\theta_1}{Z}, \ldots, \frac{\theta_K}{Z}\right)\,, \quad Z=\sum_{\ell=0}^{K_i} \theta_\ell$\\
        $\mathcal{D} \gets \mathcal{D} \cup \{ \bm{x}_i, \bm{s}_i,\bar{R}_{i}, r_{i,1},\ldots, r_{i,K} \}$ \\
        $\mathcal{P} \gets \mathcal{P} \cup \{\pi_0(\bm{s}_i|\bm{z}_i), \pi_0(s_{i,1}, 1|\bm{z}_i),\ldots,\pi_0(s_{i,K}, K|\bm{z}_i) \}$
    }
\end{algorithm}

\RestyleAlgo{ruled}
\begin{algorithm}
    \caption{Synthetic A/B test}
    \KwInput{oracle parameters $\{ \bphi, \bPsi, \bm{\gamma}, \bm{\alpha}, g_{\bm{\Gamma}}(\cdot), P_{\bm{y}}(\cdot), P_{\bm{z}}(\cdot), P_{K}(\cdot)\}$, decision rules $d_{\textsc{a}}$ and $d_{\textsc{b}}$, number of testing samples $n_{\rm test}$.}
    \KwOutput{lists of rewards $R_{\textsc{a}}$ and $R_{\textsc{b}}$.}
    $R_{\textsc{a}} \gets \{ \;\}\,, \quad R_{\textsc{b}} \gets \{ \;\}$\\
    \For{$i=1, \ldots, n_{\rm test}$}{$\bm{y}_i \sim P_{\bm{y}}(\cdot)\,, \quad \bm{z}_i \sim P_{\bm{z}}(\cdot)\,, \quad K_i \sim P_K(\cdot)$\\
        \For{$\textsc{m} \in \{\textsc{a}, \textsc{b}\}$}{
        $\bm{s}_i = (s_{i,1},\ldots, s_{i,K_i}) \gets d_{\textsc{m}}(\bm{y}_i, \bm{z}_i) \quad$ (where $d_{\textsc{m}}$ is the decision rule of $\textsc{m} \in \{\textsc{a}, \textsc{b}\}$)  \\ 
        $\theta_0 \gets \exp(\bm{y}_i^\top \bphi)$\\
        \For{$\ell=1, \ldots, K_i$}{$\theta_\ell \gets \exp (g_{\bm{\Gamma}}(\bm{z}_i)^\top \bPsi_{s_{i, \ell}}) \exp(\gamma_\ell) + \exp(\alpha_\ell)$}
        $R_{\textsc{m}} \gets R_{\textsc{m}} \cup \{1-\frac{\theta_0}{Z}\}\,, \quad Z = \sum_{\ell=1}^{K_i} \theta_\ell$
        }
    }
    \label{alg:abtest}
\end{algorithm}



\subsection{Session Completion Problems}\label{app:semi_synthetic}

We use the Twitch dataset \citep{rappaz2021recommendation} to evaluate \textbf{PRR} on user session completion tasks. For each user, we randomly split the user-item interactions $\mathcal{I}_u$ into two parts, an observed part by the baselines $\mathcal{I}_u^{\textsc{view}}$ and a hidden one $\mathcal{I}_u^{\textsc{hide}}$ that should be predicted. The task is to complete the observed user session $\mathcal{I}_u^{\textsc{view}}$ to retrieve the whole session of a user $\mathcal{I}_u$. Logged data $\mathcal{D}_n$ is collected using the \textbf{\textbf{top-K pop}} logging policy as follows. In each iteration $i \in [n]$, we randomly sample a user $u_i$. Then, we recommend a slate $\bm{s}_i$ by sampling without replacement $K$ items with probabilities proportional to their popularity. Precisely, the probability of selecting an item $a$ is $c_a/{\sum_{a^\prime} c_{a^\prime}}$ where $c_a$ is the number of occurrences of item $a$ in the dataset. After that, we construct a binary vector $b_i = (\mathbb{I}_{\{s_{i, 1} \in \mathcal{I}_{u_i}^{\textsc{hide}}\}}, \mathbb{I}_{\{s_{i, 2} \in \mathcal{I}_{u_i}^{\textsc{hide}}\}}, \ldots,\mathbb{I}_{\{s_{i, K} \in \mathcal{I}_{u_i}^{\textsc{hide}}\}}) \in \real^K$. In other words, for any $\ell \in [K]$, $b_{i,\ell}=1$ if the $\ell$-th item in the slate, $s_{i,\ell}$, is in the hidden user-item interaction $\mathcal{I}_u^{\textsc{hide}}$, and $b_{i,\ell}=0$ otherwise. This binary vector $b_i$ is then used to generate the reward and rank signals as
\begin{align}\label{eq:helper_alg}
    & \bar{R}_i, r_{i,1}, \ldots ,r_{i,K} \sim {\rm cat}\left( p_0,  p_1, \ldots, p_K \right)\,,
\end{align}
where $p_0 = \frac{\beta_0 K}{\beta_0 K + \sum_{\ell \in [K]} \beta_\ell b_{i,\ell}}$ and $p_k = \frac{\beta_k b_{i,k}}{\beta_0 K + \sum_{\ell \in [K]} \beta_\ell b_{i,\ell}}$ for $k \in [K]$. Here $\beta_0$ and $\beta_\ell$ for $\ell \in [K]$ are sampled from $\mathcal{N}(3, 9)$ and $\rm{Uniform}([16])$, respectively. This allows us to generate a dataset in the form
\begin{align*}
    \mathcal{D}_n=\{\mathcal{I}_{u_i}^{\textsc{view}}, \bm{s}_{i}, \bar{R}_i, r_{i, 1}, \ldots, r_{i, K} \, ; i \in [n]\}\,.
\end{align*}
Here $\mathcal{I}_{u_i}^{\textsc{view}}$ can be seen as the user interest features. This data generation process is summarized in \cref{alg:semi_synthetic_logs}.

\newpage

\RestyleAlgo{ruled}
\begin{algorithm}
    \caption{Session completion training logs}
    \label{alg:semi_synthetic_logs}
    \KwInput{set of $U$ users, set of $P$ items, user-item interactions $\mathcal{I}_{u}^{\textsc{view}}$ and $\mathcal{I}_{u}^{\textsc{hide}}$ for any user $u \in [U]$, number of occurrences $c_a$ for any item $a \in [P]$, maximum slate size $K$, position biases $\beta_\ell$ for $\ell \in [K]$, logging policy $\pi_0$, number of training samples $n_{\rm train}$.}
    \KwOutput{logs $\mathcal{D}$, propensity scores $\mathcal{P}$.}
    $\mathcal{D} \gets \{ \,\}\,, \quad \mathcal{P} \gets \{ \,\}$\\
     \For{$i=1, \ldots, n_{\rm train}$}{
      $u_i \sim {\rm Uniform}([U])$\\
        $K_i \sim {\rm Uniform}([K])$\\ 
        $\bm{s}_i=(s_{i,1},\ldots, s_{i,K_i}) \sim \pi_0(\cdot \mid \mathcal{I}_{u_i}^{\textsc{view}})$ \\
        $b_i = (\mathbb{I}_{\{s_{i, 1} \in \mathcal{I}_{u_i}^{\textsc{hide}}\}}, \mathbb{I}_{\{s_{i, 2} \in \mathcal{I}_{u_i}^{\textsc{hide}}\}}, \ldots,\mathbb{I}_{\{s_{i, K} \in \mathcal{I}_{u_i}^{\textsc{hide}}\}})$\\
       generate $\bar{R}_i, r_{i,1}, \ldots ,r_{i,K}$ as in \cref{eq:helper_alg}\\
        $\mathcal{D} \gets \mathcal{D} \cup \{ \mathcal{I}_{u_i}^{\textsc{view}}, \bm{s}_i,\bar{R}_{i}, r_{i,1},\ldots, r_{i,K} \}$ \\
        $\mathcal{P} \gets \mathcal{P} \cup \{\pi_0(\bm{s}_i|\mathcal{I}_{u_i}^{\textsc{view}}), \pi_0(s_{i,1}, 1|\mathcal{I}_{u_i}^{\textsc{view}}),\ldots,\pi_0(s_{i,K}, K|\mathcal{I}_{u_i}^{\textsc{view}}) \}$
    }
\end{algorithm}

After training the baselines on $\mathcal{D}_n$, they are evaluated following the data collection process except they are run instead of the logging policy $\pi_0$.

\RestyleAlgo{ruled}
\begin{algorithm}
    \caption{Session completion A/B test}
    \KwInput{set of $U$ users, set of $P$ items, user-item interactions $\mathcal{I}_{u}^{\textsc{view}}$ and $\mathcal{I}_{u}^{\textsc{hide}}$ for any user $u \in [U]$, number of occurrences $c_a$ for any item $a \in [P]$, maximum slate size $K$, position biases $\beta_\ell$ for $\ell \in [K]$, logging policy $\pi_0$, number of testing samples $n_{\rm test}$.}
    \KwOutput{lists of rewards $R_{\textsc{a}}$ and $R_{\textsc{b}}$.}
    $R_{\textsc{a}} \gets \{ \;\}\,, \quad R_{\textsc{b}} \gets \{ \;\}$\\
    \For{$i=1, \ldots, n_{\rm test}$}{
        $u_i \sim {\rm Uniform}([U])$\\
        $K_i \sim {\rm Uniform}([K])$\\ 
        \For{$\textsc{m} \in \{\textsc{a}, \textsc{b}\}$}{
        $\bm{s}_i = (s_{i,1},\ldots, s_{i,K_i}) \gets d_{\textsc{m}}(\mathcal{I}_{u_i}^{\textsc{view}}) \quad $ ($d_{\textsc{m}}$ is the decision rule of $\textsc{m} \in \{\textsc{a}, \textsc{b}\}$) \\ 
        $b_i = (\mathbb{I}_{\{s_{i, 1} \in \mathcal{I}_{u_i}^{\textsc{hide}}\}}, \mathbb{I}_{\{s_{i, 2} \in \mathcal{I}_{u_i}^{\textsc{hide}}\}}, \ldots,\mathbb{I}_{\{s_{i, K} \in \mathcal{I}_{u_i}^{\textsc{hide}}\}})$\\
        generate $\bar{R}_i, r_{i,1}, \ldots ,r_{i,K}$ as in \cref{eq:helper_alg}\\
        $R_{\textsc{m}} \gets R_{\textsc{m}} \cup \left\{\frac{\sum_{\ell \in [K]} \beta_\ell b_{i,\ell}}{\beta_0 K + \sum_{\ell \in [K]} \beta_\ell b_{i,\ell}}\right\}$
        }
    }
    \label{alg:semi_synthetic_abtest}
\end{algorithm}

\end{document}